# Crystal growth and elasticity

P.Müller

Aix-Marseille Université, Centre de Recherche sur la Matière Condensée et les Nanosciences, UPR CNRS associée aux universités Aix Marseille II et III,Campus de Luminy, case 913, F-13288 Marseille Cedex 9, France

**Abstract**

The purpose of this invited talk was to review some elasticity effects in epitaxial growth. We start by a description of the main ingredients needed to describe elasticity effects (elastic interactions, surface stress, bulk and surface elasticity, thermodynamics of stressed solids). Then we describe how bulk and surface elasticity affect growth mode and surface morphology by means of stress-driven instability. At last stress-strain evolution during crystal growth is reported.

**Introduction**

Technological development of epitaxial growth generated a new interest for studying elastic effects in crystal growth. Indeed **(1)** due to the epitaxial misfit between the deposited crystal and its foreign substrate, the deposited material is under stress and **(2)** because of the limiting sizes of new devices, stresses can reach levels which are unattainable in bulk materials. These stresses may play a role on growth mode, growth mechanism, dynamics of growth or even stability conditions. Our purpose in this invited paper is to review the main ingredients necessary to describe elastic effects in crystal growth then to illustrate their utility in some specific cases. This paper thus completes our review [1]. Obviously, since a complete treatment of elastic effects in crystal growth is still lacking, we will not use all the ingredients together but only select the more important ones for the chosen illustrations.

**1/ Ingredients:**

Elasticity effects play a role on the fundamental interactions between adatoms and the surface on which they land (section I.1) but also on the elastic properties of the further-formed phases (section I.2). At least thermodynamics of stressed solids leads to new relationships and definitions (section I.3)

      **1.1/ Elasticity effects and adatom behaviour**

Most of surface defects can be modelled by using the concept of point forces applied on a semi-infinite substrate (for a review see [2]). For example an adatom is a point defect that induces a strain field (in its underlying substrate) that can be described as the field created by an elastic dipole. On another hand, a step on a stress-free body is a linear defect that can be described by a row of elastic dipoles whereas a step on a stressed body is a linear defect whose main contribution to the elastic field can be described by a row of elastic monopoles. The elastic interactions between various kinds of surface defects are given in table I (for isotropic material and linear elasticity) where dipoles are characterized by their moment $A_{ad.}$ and $A_{step.}$ for adatom and steps respectively and where the epitaxial homogeneous strain is noted $m=\Delta a/a$.

|  | **Elastic interaction** |
|---|---|
| Between $d$-apart equivalent adatoms *(dipole-dipole)* | $U^{ad-ad}=2\dfrac{1-\nu^2}{\pi E}A_{ad}^2\dfrac{1}{d^3}$  repulsion |
| Between an adatom and a step on a body free of stress *(between a dipole and a row of dipoles)* | $U_0^{ad-step}=4\dfrac{1-\nu^2}{\pi E}A_{ad}A_{step}\dfrac{1}{d^2}$  repulsion |
| Between an adatom and a step on a stressed body *(between a dipole and a row of monopoles)* | $U_{stressed}^{ad-step}=2\dfrac{A_{ad}m}{\pi a}(1+\nu)\dfrac{a}{d}$  attraction or repulsion according to the sign of $A_{dip}m$ |
| Between two equivalent $L$-apart steps on a stressed body *(between two rows of monopoles)* | $U_0^{step-step}=2\dfrac{Ea^2m^2}{\pi}\dfrac{1+\nu}{1-\nu}\ln(L/a)$  attraction |
| Between two equivalent $L$-apart steps on a body free of stress *(between two rows of dipoles)* | $U_{stressed}^{step-step}=2\dfrac{A_{step}^2(1-\nu)^2}{\pi E a^4}(a/L)^2$  repulsion |

**Table I :** *Principal interaction laws ($E$ and $\nu$ respectively are the Young modulus and the Poisson ratio, $a$ is an atomic unit). The four last energies are given per unit length.*

Owing to the elastic interaction between an adatom and a step located at $x=x_n$, an adatom that lands on a vicinal surface is thus submitted to a force $f_n=\partial U^{ad-step}/\partial x\big|_{x_n}$ [3]. Among other effects, this elastic interaction shifts the equilibrium adatom density close to a step, which now reads for a step $n$

$$\theta_n=\theta_0\exp\left[-\frac{a^2}{kT}f_n\right] \qquad (1)$$

where $\theta_0$ the equilibrium adatom density on a step in absence of any adatom-step interaction obviously contains the elastic step-step interaction [4,5].

Let us note that strain can also affect the surface diffusion coefficient of the adatom. This effect can be analytically expressed in some situations [6,7] but remains of a second order effect in crystal growth (see section II.2).

### 1.2/ Bulk and Surface elasticity

Let us now consider elastic properties of dense phases. The bulk elastic energy change due to an infinitesimal variation of the state of deformation $d\varepsilon_{ij}$ reads :

$$dW_{elast} = \int_V \sum_{i,j} \sigma_{ij} d\varepsilon_{ij} dV \qquad (2)$$

where $\sigma_{ij}$ is the bulk stress tensor. In anisotropic elasticity $W^{elast}$ thus is a function of the epilayer growth direction. Using Hooke's law, the elastic energy has been given for a cubic system in ref. [8]: For low-index orientation it can be written as a function of the elastic constants $c_{ij}$ : $W^{elast}_{(001)} = \frac{(c_{11}+2c_{12})(c_{11}-c_{12})}{c_{11}} m^2$, $W^{elast}_{(110)} = \frac{(c_{11}+2c_{12})(c_{11}-c_{12}+6c_{44})}{2(c_{11}+c_{12}+2c_{44})} m^2$ and $W^{elast}_{(111)} = \frac{6(c_{11}+2c_{12})c_{44}}{c_{11}+2c_{12}+4c_{44}} m^2$

The expression of $W^{elast}_{(001)}$ is identical to the result obtained for isotropic material and can be written:

$$W^{elast}_{(001)} = \frac{E}{1-\nu} m^2 \qquad (3)$$

The bulk elastic energy change (2) is an extensive quantity and thus must exhibit an excess quantity when passing through the interface between two materials. For a body facing vacuum with a free surface (area $A$) normal to the $x_3$ axis, this excess energy reads [2]:

$$dW^{elast}_{Surf} = \sum_{\alpha,\beta} \tilde{s}_{\alpha\beta} A d\varepsilon_{\alpha\beta} \qquad \alpha, \beta = 1,2 \qquad (4)$$

where the quantity $\tilde{s}_{\alpha\beta}$, which defines the surface stress tensor of the given surface, is nothing else but the surface excess quantities of the bulk stress tensor components $\sigma_{\alpha\beta}$ with $\alpha, \beta = 1,2$. The surface stress tensor thus reads:

$$[\tilde{s}] = \begin{pmatrix} \tilde{s}_{11} & \tilde{s}_{12} & 0 \\ \tilde{s}_{12} & \tilde{s}_{22} & 0 \\ 0 & 0 & 0 \end{pmatrix} \text{ or } [\tilde{s}] = \begin{pmatrix} \tilde{s}_{11} & \tilde{s}_{12} \\ \tilde{s}_{12} & \tilde{s}_{22} \end{pmatrix}$$

In the second formula, surface stress is defined as a second rank tensor of row 2.

Equation (4) can be used as a definition of surface stress: <u>the surface stress is the isothermal work density done against surface deformation at constant number of surface atoms</u>. It must not be confused with the surface energy $\gamma$ which is (for a pure body) the isothermal work density done against surface creation at constant strain. Notice that Eq. (4) can be expended to upper orders in strain by introducing surface elastic constants as an excess of bulk elastic constants [2].

Because of its own surface stress, a small crystal (supposed to be cubic) has a different crystallographic parameter $a=a_0(1+\varepsilon)$ than its corresponding infinite (cubic) phase where $\varepsilon$ is the size-dependent strain of the small crystal with respect to its infinite mother phase (of crystallographic parameter $a_0$). Thus when depositing a crystal A on a semi-infinite cubic substrate B (crystallographic parameter $b_0$), the misfit between the deposited crystal and its underlying substrate is $m=(b_0-a)/a \approx m_0-\varepsilon$ where $m_0=(b_0-a_O)/a_0$ is the natural misfit (defined for the infinite phases). For 2D films (infinite lateral extension), there are two limiting cases: the pseudomorphous epitaxy for which the active misfit reads $m=m_0$ and the perfectly glissile epitaxy for which $m=-\varepsilon=-\frac{1-\nu_A}{E_A}\frac{2s_A}{h}$ ($h$ is the layer thickness and $s_A$ the surface stress of A) [9]. In the following we will essentially consider the pseudomorphous epitaxy case for which $m=m_0$.

### 1.3/ Thermodynamics of stressed solids

Another important ingredient necessary for describing thermodynamic transitions (as crystal growth) is the concept of chemical potential. However, for crystalline solids under anisotropic stress, this concept is far from being evident as first evidenced by Gibbs [10], discussed by Rusanov [11] but really rationalised by Cahn and Larché [12]. Let us consider both approaches

\* When Gibbs wrote his famous paper [10], solid-state diffusion was unknown so that he described a solid as a body composed of immobile components (that cannot diffuse in the solid) possibly in equilibrium with internal mobile components (a fluid that may diffuse and even distort the solid[1]). Gibbs avoided defining any chemical potential for immobile species, but considered the equilibrium of such a solid with an external fluid. The <u>equilibrium condition</u> he found, states that the chemical potential $\mu_{j,\alpha}^{fluid}$ of the external fluid species $j$ in contact with the solid face $\alpha$ normal to the direction $\vec{x}_\alpha$ (with $\alpha=1,2,3$) depends upon the surface orientation[2] and reads [10,11]:

$$\left(f-\sigma_{\alpha\alpha}-\sum_i \mu_i^{sol}\rho_i\right)\bigg/\omega_j=\mu_{j,\alpha}^{fluid} \qquad (5)$$

---

[1] For instance think about a foreign gas dissolved in a metal, the electron gas in a metal or even more exotic body as a crystalline sponge filled with a fluid.
[2] It is not true if one allows solid component to diffuse by vacancies creation, exchange and annihilation [12]

where $f$ is the free energy density of the solid, $\omega_j$ the molecular volume of the species $j$ and $\mu_i^{sol}$ the chemical potential of the mobile species $i$ (of concentration $\rho_i$) in the solid. An anisotropic solid can thus be in chemical equilibrium with three liquids (facing *e.g.* the opposite faces of an elemental cube) at different pressures $\sigma_{\alpha\alpha}=-P_\alpha$. Notice that whereas Gibbs avoided defining a chemical potential for immobile species (they cannot diffuse !), other authors [11] used (5) to introduce the concept of <u>chemical potential tensor</u> valid for immobile species, the principal values of which read $\mu_{j,\alpha\alpha}^{sol}=\mu_{j,\alpha}^{fluid}$ when in equilibrium with a fluid facing the solid face $\alpha$.

* In the 70$^{th}$-80$^{th}$, Larché and Cahn [12] studied the internal equilibrium of strained crystals in which <u>solid-state diffusion is active</u>. For this purpose they introduced a crystalline network restriction: an atom located on a network site cannot move freely but can only be exchanged among crystalline sites. Furthermore they distinguished the possible diffusion of substitutional and/or intersticial species. Whereas for a fluid that contains $N$ components, the condition of mass conservation usually leads to $N$ Lagrange multipliers $\mu_j$ (the chemical potentials) <u>constant in the whole fluid volume</u>, for a crystalline solid, the additional network restriction leads, for substitutional species $k$, to only $N-1$ Lagrange multipliers $M_{k,K}=\partial u/\partial c_k - \partial u/\partial c_K$ ($u$ is the internal energy density) <u>constant in the whole crystal volume</u> [3]. Since describing the exchange of two atomic species $k \leftrightarrow K$, $M_{k,K}$ coefficients are called diffusion potentials[4]. The main result is the coupling between chemical composition and stress: the local equilibrium composition is modified by stress (and vice versa) [12].

The free energy density change thus reads, if there are $L$ interstitial species and $N-L$ substitutional ones, ($s$ is the entropy density):

$$df = -sdT + \sum_{\alpha,\beta}\sigma_{\alpha\beta}d\varepsilon_{\alpha\beta} + \sum_{k=L+1}^{N-1}M_{k,K}d\rho_k + \sum_{k=1}^{L}\mu_k d\rho_k \qquad \text{for the bulk} \qquad (6a)$$

$$df^{Surf} = -s^{surf}dT + \tilde{s}_{\alpha\beta}d\varepsilon_{\alpha\beta} + \sum_{k=L+1}^{N-1}M_{k,K}d\rho_k^{surf} + \sum_{k=1}^{L}\mu_k d\rho_k^{surf} \qquad \text{for a free surface} \qquad (6b)$$

where in (6b) we used (4) and where the subscripts *surf* indicate surface excess quantities. Notice that: **(i)** in absence of any interstitial species the last term of equations (6) vanishes and **(ii)** for a crystal in equilibrium with a fluid $M_{k,K}$ simply reads $M_{k,K}=\mu_k^{fluid}-\mu_K^{fluid}$, so that if vacancies are chosen as the dependent $K$ species, there is $M_{k,Vac.}=\mu_k^{fluid}$ [12]. Fundamental thermodynamics equations, as well as equilibrium conditions valid for various situations can be found in [12].

---

[3] Again, no definition of individual chemical potential of substitutional species arises in the solid.

[4] Obviously, a chemical potential $\mu_i$, constant in the whole volume, can still be attributed to the intersticial species $i$ since they do not obey to any network restriction

* Last but no least, notice that more complete developments of surface thermodynamics allow defining the Shuttleworth relation (7) connecting the surface stress tensor $\tilde{s}$ to the surface energy $\gamma$, of a given body, via the derivative of the surface energy with respect to the bulk strain extrapolated to the surface [2,13,14]:

$$\tilde{s}_{\alpha\beta}-\gamma\delta_{\alpha\beta}=\frac{\partial \gamma}{\partial \varepsilon_{\alpha\beta}}\bigg|_{T,\mu} \qquad (7)$$

It is thus clear that, since it is not possible to deform a fluid, surface stress and surface energy of a fluid cannot be distinguished.

**2/ Growth mode and elasticity:**

**2.1/ Growth conditions**

The most important ingredients necessary to describe the growth modes close to equilibrium are the wetting conditions and the super-saturation conditions. As we will see in the two following sections elasticity plays a role on both ingredients.

* From an experimental viewpoint, three growth modes of a crystal A onto a crystal B have been recognized: the 3D (or Volmer-Weber mode), the 2D (or Frank van der Merwe growth) and the mixed 2D followed by 3D mode (Stranski-Krastanov mode). In absence of elasticity, Bauer [15] rationalized these growth modes by defining the so-called wetting factor $\Phi = 2\gamma_A - \beta = \gamma_A + \gamma_{AB} - \gamma_B$ where $\gamma_n$ is the surface energy of the material $n$, $\gamma_{AB}$ the interfacial energy and $\beta$ the adhesion energy. Then for energetic reasons, 2D growth occurs for $\Phi < 0$ while 3D growth occurs for $\Phi > 0$. Obviously, in presence of elasticity the surface stress must intervene. There are many ways to obtain the surface stress effect on the wetting condition via a thermodynamic process or a development of the wetting factor $\Phi$ with respect to the deformation [9,16] followed by the use of relation (7) connecting the surface energy derivative with respect to strain to the difference between surface energy and surface stress (in a Eulerian picture). For 2D pseudomorphous epitaxy, there is (when neglecting any elastic relaxation in the film[5])

---

[5] In case of 3D pseudomorphous epitaxy, (but still in the thin film approximation so that there is no normal interfacial stress) the strain relaxation of the deposit has to be taken into account so that (8) becomes: $\Phi = 2\gamma_A - \beta + 2\varepsilon^{top}(\tilde{s}_A - \gamma_A) + 2m_0(\tilde{s}_A - \gamma_A)$ where $\varepsilon^{top} < m_0$ is the (size dependent) relaxed strain in the upper layer. If moreover the underlying substrate is dragged during the elastic relaxation there must be an

$$\Phi = 2\gamma_A - \beta + 4m_0(\tilde{s}_A - \gamma_A) \qquad (8)$$

The growth mode thus can be changed by elasticity. The change is all the more important the misfit $m$ is high and/or the difference between surface stress and surface energy is high (for a quantitative discussion see [16]).

   \* Another criterion necessary to define crystal growth conditions is the supersaturation condition $\Delta\mu$ defined as the chemical potential difference between the crystal and its mother phase[6]. It is well known that in absence of elasticity, 3D growth takes place at supersaturation ($\Delta\mu>0$) while 2D growth can occur at undersaturation ($\Delta\mu<0$) [17]. In presence of elasticity, the criterion becomes (for isotropic solids):

$$\Delta\mu < \frac{E_A}{1-\nu_A^2} m_0^2 \quad \text{with } \Phi<0 \quad \text{for 2D growth} \qquad (9a)$$

$$\Delta\mu > \frac{E_A}{1-\nu_A^2} m_0^2 R^{shape} \quad \text{with } \Phi>0 \quad \text{for 3D growth} \qquad (9b)$$

The factor $0 < R^{shape} \leq 1$ in (9b) describes the elastic relaxation of the 3D crystal. It is unity for unrelaxed crystals. Relation (9a) states that 2D growth occurs at undersaturation with respect to the elastic energy density stored by the 2D layers, whereas relation (9b) states that supersaturation with respect to the elastic energy density of the <u>relaxed</u> system is necessary for 3D growth. Obviously since the 3D crystal relax by its free edges, $R^{shape}$ depends on the size and the shape of the deposited crystal. Moreover, since the deposit may also, during its elastic relaxation, deform its underlying substrate $R^{shape}$ also depends upon the relative stiffness of the substrate and its deposit. $R^{shape}$ has been calculated in some specific geometries, then used to calculate the 3D equilibrium shape of the partially relaxed deposited crystal (see section IV) [18-21].

Notice that previous conditions (9) open a window for Stranski Krastanov (SK) growth close to equilibrium conditions [22]:

$$\frac{E_A}{1-\nu_A^2} m^2 R^{shape} < \Delta\mu < \frac{E_A}{1-\nu_A^2} m^2 \quad \text{with } \Phi<0 \quad \text{for 3D/2D growth} \qquad (9c)$$

But, SK growth can also proceed from non-equilibrium mechanisms.

---

additional work necessary to deform the interface from $m_0$ to $\varepsilon^{int}$. Expression (8), exact for 2D films (no elastic relaxation), overestimates the elasticity effect in case of 3D growth.

[6] During crystal growth from a fluid phase, atomic species are supposed to be incorporated via the crystal surface and not to diffuse in the solid state so that $\Delta\mu$ is unambiguously defined (see section I.5).

## 2.2/ Growth mechanisms

All the elastic ingredients we describe also play a role on the growth mechanisms. Let us leave aside the (obvious) fact that inhomogeneous strain can create preferential sites for 2D or 3D nucleation but let us underline elastic effects on growth mechanisms we will partially describe in the following sections. <u>For 2D growth,</u> **(i)** since the adatom density essentially depends upon the ratio $D/F$ (surface diffusion constant $D$, incoming flux $F$), a change of $D$ can shift the crossover from step flow to 2D nucleation (or the reverse), **(ii)** the elastic repulsion between steps can promote step flow rather than step bunching (see below). <u>For 3D growth</u>, elasticity modifies equilibrium and growth shapes. In both cases (2D and 3D), the elastic relaxation that lowers the total energy can give birth to new types of morphological instabilities described in the following section. At last <u>SK transition</u> is completely governed by elastic relaxation [16,22]

## 3/ Instabilities of thin films

As soon as the surface of a 2D homogeneously strained film is perturbed, the stresses are no longer constant but can relax at hills and concentrate in valleys. A strained film can thus be unstable to the formation of surface "bumps", the nature of which depends on the surface state (flat, stepped or rough). In the following, we will focus on thermodynamics effects in absence of surface stress. The effect of the impinging flux (kinetics effects) will be only mentioned.

\* For epitaxial films with a rough surface (at the atomic scale) the morphological change can be <u>continuously</u> described by a hill and valley structure. The simpler case concerns thick films whose behaviour is thus similar to the one described for semi-infinite strained solids [23-25]: The instability is governed by the balance between the surface stiffness $\tilde{\gamma}=\gamma+\partial^2\gamma/\partial\theta^2$ of the film[7] (stabilizing term) and the elastic energy (destabilizing term). The planar surface is thus unstable for perturbations with wavelengths greater than the critical value $\lambda_{crit} = \pi \frac{\tilde{\gamma}}{E_A m_0^2} \frac{1-\nu_A}{1+\nu_A}$ [23-25]. For thin film deposited on a foreign substrate, things are more complex since the wetting interaction between the film and the underlying substrate as well as the substrate rigidity also plays a role [26,27]. Considering $\Phi=0$, Spencer et al [27] show that a flat film on a rigid substrate is stable against small perturbations of all wavelengths provided its thickness $h$ is smaller than some critical value $h_c \propto \tilde{\gamma}/m_0^2$. For

---

[7] $\theta$ defines the local orientation of the surface, but notice that most authors mix $\gamma$ and $\tilde{\gamma}$.

non-rigid substrates no such critical thickness exists. In other words, for $\Phi=0$, deposited films are always instable excepted when the substrate is perfectly rigid. However obviously wetting interactions $(\Phi<0)$ as well as surface stress can act to stabilise again the film [26]. Notice, that the non-linear evolution of the instability has been largely studied and leads to the formation of deep cusps [28,29]. For a film growing on a rigid substrate, the critical thickness depends on the growth rate $F$ and becomes $h_c \propto F\gamma^3/m_0^8$ [27]

* For vicinal (stepped) surfaces, the morphological instability is driven by step-bunching mechanism [3-5]. The main ingredient that allows describing this instability is the shift (Eq. (1)) of the equilibrium adatom density induced by the elastic interactions. The critical wavelength beyond which there is a linear instability (in the regime of very strong repulsion) reads $\lambda_{crit} = 2\pi^2 \frac{A_{step}^2}{L} \frac{(1-\nu)^2}{E^2 m_0^2}$ where $L$ is the inter-step distance [5]. Thus, it is now the force dipole moment density (describing the inter-step interaction), and no more $\tilde{\gamma}$, which is the stabilizing factor (the equilibrium state now is an equilibrium of steps). In presence of an external flux, two limiting cases can be easily discussed [5]. In the limit of strong flux, the planar surface is unstable only when adatom incorporation is privileged on the lower step. In the limit of weak flux, but without any Schwoebel barrier, a long-wave instability is recovered [5]. Notice that stress also alters usual step meandering instabilities. It plays no role in the determination of the most stable mode but can drastically affect the development of the instability [30]. Furthermore, bunching and undulation mechanisms compete each other: for small step spacing, step bunching dominates while for large step spacing meandering dominates. The crossover from bunching to undulations is exponentially sensitive to the stress [31].

* For flat surfaces, the bump formation proceeds by islands and holes nucleation [31,32]. The equilibrium morphology can thus consist of 3D islands sit on a thin wetting layer (Stranski-Krastanov growth). The morphology of the islands is governed by the balance between surface energy and elastic energy while the thickness of the wetting sublayer is governed by the balance between the elastic energy and the wetting potential [26,34].

**4/ Stress-strain evolution during island growth:**

Let us consider separately 2D, 3D then self-assembled growth.

* In-plane lattice oscillations associated to 2D pseudomorphic growth regime have been reported [35]. This behaviour originates from the fact that 2D strained islands may elastically relax by their free edges. As a consequence, during perfect lateral growth of 2D

islands, the mean in-plane lattice spacing oscillates with a maximum (minimum) for half (complete) coverage. For homoepitaxially system, the oscillation originates from the surface stress of the 2D islands which allows to define an active misfit [36] $m=0-\varepsilon$ where $\varepsilon$ depends on the size of the 2D islands[8] (see section 1.2).

* 3D epitaxial deposits when accommodated on a mismatched substrate only reach an equilibrium state for a given shape and a given strain distribution (in the deposit and its substrate). The equilibrium shape of the 3D islands thus is no more the shape that minimizes the surface energy, but the one that minimizes the total energy. Elastic relaxation takes place by the island edges and thus depends on the crystal shape (see eq. (9b) and associated comments). As a consequence, the equilibrium shape becomes size-dependent and thus changes during crystal growth close to equilibrium conditions. More precisely since the epitaxial stress acts against wetting, it globally leads to a thickening of the equilibrium shape as largely discussed in [19,37]. Though we are not concerned with plasticity, notice that for great-enough crystal sizes, dislocation entrance may relax abruptly the stress. The equilibrium shape thus exhibits a saw-tooth behaviour with alternatively thickening (between two dislocations entrance) then flattening (at each dislocation entrance) of the crystal shape [18,37].

* Since the elastic deformation the islands induce in their underlying substrate can overlap, the islands communicate via the substrate. This overlapping induces some interesting phenomena. Indeed if, in an array of islands one of the islands deviates in size or shape its neighbours islands feel the change so that a driving force for restoring a uniform size and shape distribution appears. At the same, if the position of an island, in a periodic array, deviates from its lattice site, the overlapping of the strain-stress field inside the crystal becomes asymmetric which is at the origin of a new driving force which restores the identical distances as analytically shown in [18]. Recent kinetic Monte Carlo simulations [36] allow modelling the growth of strained semiconductors. It is thus unambiguously shown that the strain field creates spatial ordering and narrows the size distribution. It is even the case for step flow growth in multilayers films, for which the inter-layer step-step interaction can lead to vertical self-organization [39,40].

---

[8] $\varepsilon$ can be written as a function of the surface stress of the material A (mesoscopic approach) or with pair potential (atomistic approach) [36].

## 5/ Conclusion

We focalised on elasticity effects in crystal growth and thus to elastic distortion and elastic-induced roughening. However, we have neglected two other mechanisms that can play an important role: strain relaxation by introduction of misfit dislocations as well as interdiffusion. For 2D layers, the maximum thickness above which it is not possible to grow pseudomorphic layers without introducing dislocations (the so-called critical thickness) has been largely studied (for a review see [41]) . Such dislocations can even be at the origin of technological applications when the stress field they induce is used for selecting the nucleation sites [42]. The interplay between interdiffusion and strain relaxation mechanisms has essentially been studied in case of Si/Ge growth [43].

These plastic mechanisms can be a severe limitation to what precede. Indeed, at high temperature where atomic diffusion and defect creation are thermally activated, plasticity effects take over and elasticity may become ineffective. It is for this reason that the most beautiful experiments that put in evidence the stress-induced morphological instabilities have been performed with He crystals that means with a defect-less crystal at low temperature [44].

**Acknowledgements:**

I would like to acknowledge R.Kern, F.Leroy, J.J. Métois and A.Saùl for very helpful discussions